# Measuring Open Access Publications: A Novel Normalized Open Access Indicator

*Abdelghani Maddi*

*Observatoire des Sciences et Techniques, Hcéres,
2 rue Albert Einstein, Paris, 75013 France*

*abdelghani.maddi@hceres.fr
ORCID: https://orcid.org/0000-0001-9268-8022*

**Abstract**
The issue of Open Access (OA) to scientific publications is attracting growing interest within the scientific community and among policy makers. Open access indicators are being calculated. In its 2019 ranking, the "Centre for Science and Technology Studies" (CWTS) provides the number and the share of OA publications per institution. This gives an idea of the degree of openness of institutions. However, not taking into account the disciplinary specificities and the specialization of institutions makes comparisons based on the shares of OA publications biased. We show that OA publishing practices vary considerably according to discipline. As a result, we propose two methods to normalize OA share; by WoS subject categories and by disciplines. Normalized Open Access Indicator (NOAI) corrects for disciplinary composition and allows a better comparability of institutions or countries.

**Article Highlights**
1. Make an inventory of Open Access, by OST disciplines and ERC sub-fields, in the Web of Science database between 2000 and 2017.
2. Propose a novel normalized Open Access indicator for a better comparability of different actors.
3. Provide a state of the art in Open Access publishing.

**Classification codes**

**Acknowledgements**
The author would like to thank Frédérique Sachwald, Lesya Baudoin and Mathieu Goudard for their comments on an earlier version and for their guidance in improving the manuscript. The author also wishes to thank the referees for their comments and suggestions, which have contributed to improve the paper.

## 1. Introduction

The issue of open access (OA) to publications is attracting growing interest within the scientific community and among policy makers. OA reduces barriers to accessing research results, which represents a better dissemination of knowledge and contributes to the development of science (Martín-Martín *et al.* 2018). The scientific literature on this issue shows that OA publications are much more cited than their counterparts for whom no Open Access version is available (Antelman, 2004; Harnad *et al.* 2004; Eysenbach, 2006; Piwowar *et al.* 2018). Thus, the academic impact of researchers and institutions increases as the number of OA publications increases. As a result, researchers are increasingly led to publish in OA in order to make their results more accessible, with the prospect of a higher and a faster impact (Antelman, 2017).

For funders, the stakes are different as OA does not necessarily mean "free" and may even generate new costs (Borrego, 2016; Anderson, 2017a, 2017b). In addition to the subscription costs that institutions are enduring granting their researchers access to publications, they are now more and more led to pay the costs of publications in OA. These costs can reach 5000 euros for one publication (Simth *et al.*, 2017; Antelman, 2017). This amounts to pay twice for OA publications. For this reason, some consider that the current system based on publisher subscriptions becomes anachronistic and it is imperative to upgrade in to 100% OA. The concept of "Big Deal" emerged to denote the difficulty of changing the publishing market system as it works nowadays (Björk, 2016b; Schiermeier & Mega, 2017; Anderson, 2017a; Université Konstanz, 2014; Université de Montréal, 2017). Schimmer et al. (2015) showed that if the Web of Science (WoS) only indexed articles (1.5 million in 2013), the unit cost in the current subscription system would be around 5000 euros per article (the overall cost of subscriptions is estimated at EUR 7.6 billion). While in a system that only operates according to OA rules, the community would produce 2 million articles at a unit cost of 3800 euros (with the same budget) (Schimmer et al. 2015).Several countries, like the Netherlands, Germany and the United Kingdom have started negotiations with publishers to find an agreement around the "Big Deal" including subscriptions and Article Processing Charges (APCs). In France, after 13 months of negotiations, in order to limit the rise in subscription prices and take into account APCs, the national consortium has decided not to renew the agreement with Springer since 2018[1].

Over the last ten years, the scientific community has witnessed a rise in a discourse aligned with the funders' perspective. A large scientific community agrees that research results should be accessible not only for all researchers but also for society as a whole (Tennant *et al.* 2016). Since research is funded mainly by taxpayers, it is unjustified that publications are held exclusively by multinationals, which are demanding increasingly high fees (Subscription and APCs fees). In addition, sharing research content immediately, opening up science also has virtues at the global level. Due to lack of funds, some researchers in low-income countries do not have the same access to publications as their counterparts in high-income countries. Moving to a 100% OA system would provide greater equity (Schöpfel, 2017).

In this context, public policy makers and research funding institutions have set targets regarding open access and want to be able to track changes in relevant indicators. As a result, they become the first seekers for OA indicators.

---

[1] https://www.couperin.org/services-et-prospective/grilles-d-evaluation-ressources/261-a-la-une/1333-couperin-ne-renouvelle-pas-l-accord-national-passe-avec-springer

Measuring the degree of commitment in the open science movement of a given actor can, for example, be approached in a simple way by calculating its share of open access publications. Some centers specializing in the production of science indicators now include OA indicators per institution. In its 2019 ranking, the "Centre for Science and Technology Studies" (CWTS, 2019: https://www.leidenranking.com/ranking/2019/list) provides open access indicators (numbers and share) of institutions with all variations by OA type (the methodology of OA indicators in the CWTS ranking is detailed in: Van Leeuwen et al. 2019). This gives an idea about the degree of openness of institutions.

However, not taking into account the disciplinary specificities and the specialization of the institutions makes comparisons based on the OA share biased. Thus, open access publishing practices vary considerably by discipline. OA share is very high in Fundamental Biology and much less so in Computer Science and Engineering (see European Commission 2019; Gargouri et al. 2012; Kozak et al. 2013; Zhu 2017). In addition, the estimates of OA shares can vary considerably depending on the sources used (full integration of the archives or only the articles published in journals) and of the perimeters considered (for example, taking into account or not publications without Digital Object Identifier (DOI)). For the case of France, the estimate of the OA share is 41% in the study of the European Commission (2019) (Scopus and Unpaywall), against 30% in the Observatoire des Sciences et Techniques (OST) calculations (WoS).

Making comparisons based on simple OA shares can be hazardous. It is customary in the scientific community that institutions compare themselves by simply using the share of open access publications. To content with comparing only the OA shares biases the judgment drawn from the degree of openness of a given institution. This amounts to ignoring the environment in which it operates, driven by globally shared publishing practices. It is therefore imperative to consider an actor's OA share relatively to the overall share of OA worldwide, in order to have a comparable indicator regardless of the actor. A University Hospital Center will systematically have a larger OA share than an institution with a strong IT component. Including all institutions, with their diversity, in the same comparison can be risky in terms of public policy.

The objective of this paper is twofold. First, draw up an overview of the OA publications in the WoS database in terms of volume, evolution and disciplinary distribution. Second, propose two methods of normalization of OA share; by WoS subject categories and by OST disciplines. This indicator corrects OA share by taking into account disciplinary practices. This allows a better comparability of different actors (institutions and countries). To the best of our knowledge, there is no study that uses this type of normalization to compute Open Access indicators.

The rest of the article is organized as follows. Section 2 presents the data and the normalization method. Section 3 presents some descriptive open access statistics based on OST in house WoS database. Section 4 applies the standardization method at the country and institution levels. Finally, we discuss in the conclusion the factors that can affect the numbers and open access shares and the importance of normalization.

## 2. Data & method

### 2.1. OST database and open access data

The data has been extracted from the Observatoire des sciences et Techniques' (OST) in-house database. It includes five indexes of the Web of Science (WoS) available from Clarivate Analytics (Science Citation Index Expanded (SCIE), Social Sciences Citation Index (SSCI), Arts & Humanities Citation Index (AHCI), Conference Proceedings Citation Index (CPCI-SSH) and Conference Proceedings Citation Index (CPCI-S)) and corresponds to WoS content indexed through the end of March 2019.

The choice to work only on the WoS database offers the possibility of working on a fine, stable and validated classification (254 WoS subject categories[2]) which serves as the basis for normalization. The OST has also implemented two more aggregated classifications. The first contains 11 disciplinary levels (see annex 2). The second is based on the European Research Council (ERC) panels (see https://erc.europa.eu/content/erc-panel-structure-2019). The advantage of this nomenclature is that it has a European and an international dimension and contains two levels of aggregation: one level in three large fields (LS: Life Sciences, PE: Physical Sciences and Engineering and SH: Social Sciences and Humanities) and one level in 25 sub-fields (9 in LS, 10 in PE and 6 in SH). The OST database is therefore ready for in-depth characterization studies of scientific publications.

Since 2014, the provider of the WoS database, Clarivate Analytics (CA), retrospectively identifies the status of OA publications. In 2017, CA signed a partnership with *ImpactStory* to better identify OA status. More recently, at the end of 2019, *ImpactStory* changed its name to *Our Research* (https://our-research.org/). *Our Research* now uses the *Unpaywall* database to identify the open access status of publications.

There are two main types of OA publications; "Gold" and "Green" (Björk *et al.* 2010; Björk *et al.* 2014; Björk, 2017). Both of types allow readers to access into the full text. Gold OA covers especially Creative Commons licensed articles published in journals listed in the Directory of Open Access Journals (DOAJ) (Gargouri *et al.,* 2012; Archambault *et al.,* 2014; Bolick, 2017). These are journals that rely on an economic model based on APCs paid by authors (usually via their institution). Green OA represents articles deposited in open archives. Some non-OA journals allow authors to submit either the version before the peer review (*Preprint*) (see Guédon, 2004) or the post-evaluation – peer reviewed – version (*Post-print*). Apart from these two types, there is another category of OA called "Bronze". It includes articles published in journals that do not have a license (Creative Commons) or articles with an unidentified OA status (which can be temporary) in databases. Their status can evolve over time to become Green OA.

It is important to distinguish between the status of publication and that of journal. A journal can have three statuses; OA, not OA or hybrid. An OA journal publishes OA-type articles, while an hybrid journal is a closed (fee-paying) journal that gives authors the option to publish in OA for a fee (APCs). The resulting publication will also be of the "Gold" OA type (Walker *et al.* 1998; Laakso *et al.* 2012, 2013; Björk, 2016a; Martín-Martín *et al.* 2018). It is possible for an article to have multiple OA statuses at the same time. For example, an article can be published in an OA journal (Gold) and then deposited in an archive (Green). More generally, beyond the deposits made by the authors, OA journals can fully deposit their contents in an archive like *PubMed Central*. It should be noted that the reliability of the information varies: the information on Gold OA can be considered reliable and stable, while the Bronze status is volatile by nature. Open archives are fed continuously by the authors, their institutions or publishers, the information on Green OA is thus also quite ephemeral (Björk, 2016a; Martín-Martín *et al.* 2018).

For publications with multiple open access status, we have established an order of priority, as follows: gold, bronze then green. For example, a publication that is both Gold and Green is considered in our calculations as Gold only. The same reasoning is applied for all multi-status publications in the order indicated. The order is chosen according to an institutional logic. Gold generally results from the initiative of institutions that pay for APCs; Bronze is on the initiative of journals (editors) and finally Green at the initiative of authors.

---
[2] See: https://images.webofknowledge.com/images/help/WOS/hp_research_areas_easca.html

### 2.2. IPERU program and Institutional data

OST IPERU program ("University Research Institutions Output Indicators") provides a set of bibliometric indicators to 126 French universitary institutions.

IPERU indicators are used to describe the scientific and the technological output of the institutions, to monitor their development and to assess their positioning in reference geographical areas. For more details about the program, (see: https://www.hceres.fr/en/iperu-programme).

Of the 126 institutions, we selected for the study only those that contain a number of publications greater than 30. That makes a total of 112 institutions. The OST proceeds to the disambiguation and the unification of the addresses and affiliations, thus enriching the WoS data.

The 112 institutions are divided into 3 groups according to their number of publications and their disciplinary orientation (see Table 1).

**Table 1: Description and number of French universitary institutions per IPERU group**

| Group | Number of institutions | Description |
|---|---|---|
| G1 | 54 | Includes large universities and research organizations (with at least 500 publications per year). The majority of institutions in this group contain at least one component in science or medicine. Of the 54 institutions, there are 11 major engineering schools. |
| G2 | 37 | Includes Social Sciences and Humanities (SSH) universities, engineering schools and multidisciplinary institutions, most of which do not have a medical component. The average number of publications in this category is between 150 and 500 per year. |
| G3 | 21 | Includes small engineering schools and some small universities or SSH universities with less than 150 publications a year. |

### 2.3. Normalization method

We calculate the Normalized Open Access Indicator (NOAI) for the first 50 producing countries and for the French institutions included in the 2019 IPERU-program.

We apply disciplinary fractional counting to compute the publications counts. The disciplinary fractional counting involves dividing the credit by the number of fields (subject category or OST discipline) to which a publication is assigned, while keeping whole counting for the geographical dimension. Whole counting involves assigning full credit to each signatory of a given publication. Table 2 illustrates the method of counting used with the example of a publication co-signed by two authors affiliated with two institutions, one in France and the other in The Netherlands. This publication is also assigned to three different WoS categories and two disciplines.

**Table 2: Counting method for multidisciplinary publications**

|  |  | Country fraction (whole count) ||
|---|---|---|---|
| Field fraction | Discipline fraction | France:1 | The Netherlands: 1 |
| Medical Informatics: **0.33** | Computer science: **0.66** | 0.33*1 = 0.33 | 0.33*1 = 0.33 |
| Computer Science, Information Systems: **0.33** |  | 0.33*1 = 0.33 | 0.33*1 = 0.33 |
| Health Care Sciences & Services: **0.33** | Medical research: 0.33 | 0.33*1 = 0.33 | 0.33*1 = 0.33 |

While it also possible to consider a fractional counting on the geographical dimension, and a combined fractional counting on both dimensions, we decide to apply only the disciplinary fractional counting since geographical fractionating does not seem to us to conceptually make much sense for open access indicators.

Normalization is applied in two stages. First, calculate the share of OA publications ($OA_{ij}/x_{ij}$) by institution (or country) and by discipline (or field – WoS subject category), then normalize by the world share ($OA_{wj}/X_{wj}$).

$$OA_{S_{ij}} = \frac{OA_{ij}/x_{ij}}{OA_{wj}/X_{wj}}$$

In a second step, to have an overall OA indicator by institution (or country), it is possible to calculate a weighted average by the number of publications per discipline. We obtain then the Normalized Open Access Index (NOAI):

$$NOAI_i = \frac{\sum(OA_{S_{ij}} \times x_{ij})}{x_i}$$

### 3. Open access by discipline and country

The WoS Core Collection (all indexes) has 5,000 journals in open access out of a total of 26,400 journals (without hybrid journals). The number of hybrid journals is not given in the statistics provided by Clarivate Analytics. Out of a total of 75 million publications, the WoS Core Collection database provides open access status for 12 million (see: https://clarivate.com/webofsciencegroup/solutions/open-access/ accessed February 26, 2020).

The statistics obtained from the OST database are different. They are calculated for four types of publication (articles, letters, reviews and conference proceedings) and four indexes (see section 2.1.). In March 2019, the OST database had 27,500,000 publications, of which 6,420,000 had an open access status. OST database contains 14,000 journals, of which 2,000 are open (100% with APCs or funded by institutions).

**Figure 1: share of world Open Access publications (WoS)**

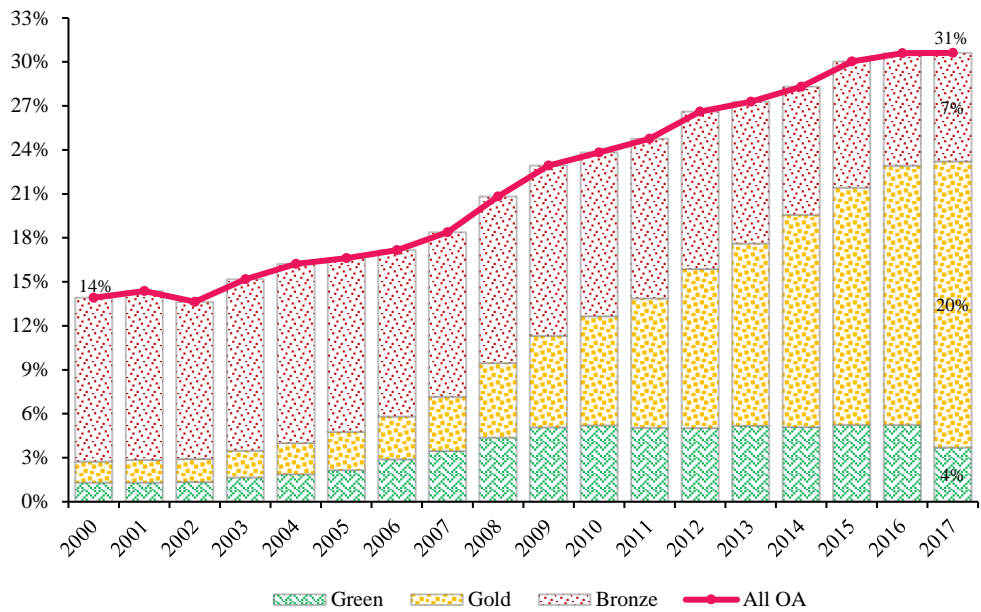

Source: Computed by author using OST-WoS database

Figure 1 shows that the share of OA publications has increased by more than 100% between 2000 and 2017, reaching 31% at the end of the period. The bulk (about 2/3) is "Gold" publications (or "Bronze" which is gradually transformed especially into "Green"). This distribution is a peculiarity of WoS and does not necessarily reflect the real practices of "openness" in these disciplines. For example, OA share would be much higher in mathematics and physics if open archives such as ArXiv were *fully* taken into account. Indeed, the green OA status found in international databases such as WoS, only concerns books or articles published in journals, in conference proceedings or in books series. Not to be confused with invisible green open access in databases, which notably concerns articles not published in journals. For example, an article in physics simply deposited in an open archive will be considered in the WoS data only if it has been published in a journal indexed in the database. The "Green" status in databases only concerns articles deposited in archives AND published in indexed journals.

This shows that it is imperative to take into account the representativeness of the database used for this indicator and that it is essential to normalize when comparing research actors (given their disciplinary orientations).

The share of Gold OA has increased very significantly over the period, from 1.4% in 2000 to 20% in 2017. Green Open access also increased, but to a lesser extent, 1.4% in 2000 to around 5% after 2009. By contrast, the share of bronze OA fell over time. It goes from 11% to 7%. This is due in part to the end of embargo period for non-open publications. After the embargo, certain journals self-archive publications, which therefore become Green OA.

**Figure 2: share of world OA publications by ERC sub-fields**

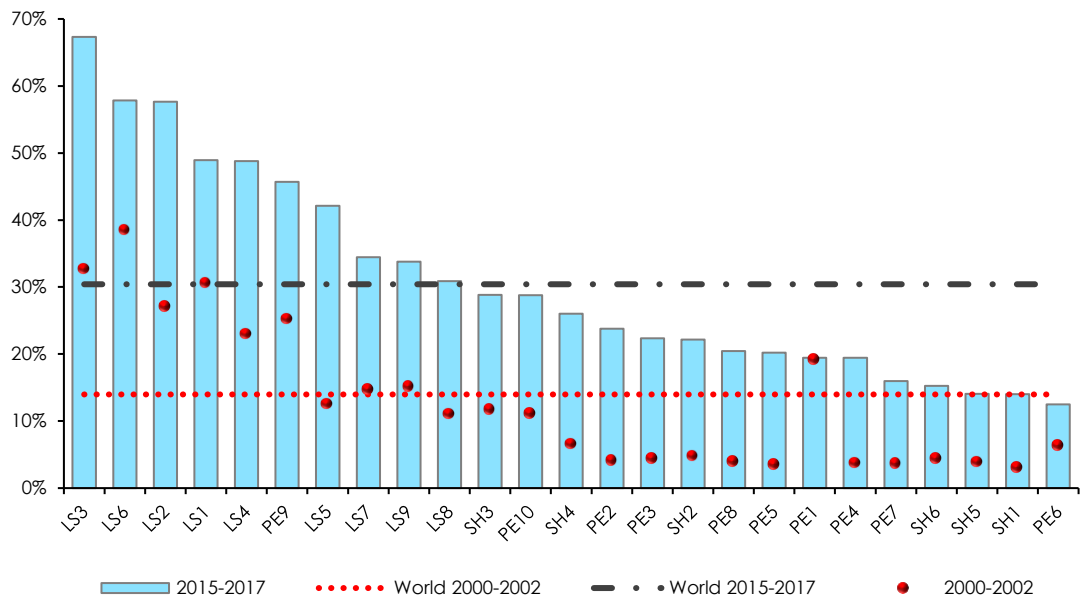

Source: Computed by author using OST-WoS database

Figure 2 shows the distribution of OA publications according to the European Research Council (ERC) subfields (for labels, see annex 3). There is a great disparity between the different sub-fields as to the "practices of openness". The share of OA publications has increased significantly for almost all sub-fields. The "Mathematics" sub-field (PE1) is the only one to remain stable over the period. The explanation of this stagnation requires an in depth study.

In 2015-2017, the share varies between 12% for the Computer Science and Informatics (PE6) sub-field and 70% for the Cellular and Developmental Biology (LS3) sub-field. Overall, the share of OA is relatively high in the areas of "Life Sciences" (LS), and low in particular in the sub-fields of "Social Sciences and Humanities" (SH) and "Physical Sciences and Engineering" (PE). Leaving aside the Universe Sciences (PE9) sub-field with a 46% proportion of OA the remaining sub-fields of PE and SH domains have a lower than world average share in both periods (2000-2002 and 2015-2017).

**Figure 3: share of world OA publications by OST disciplines**

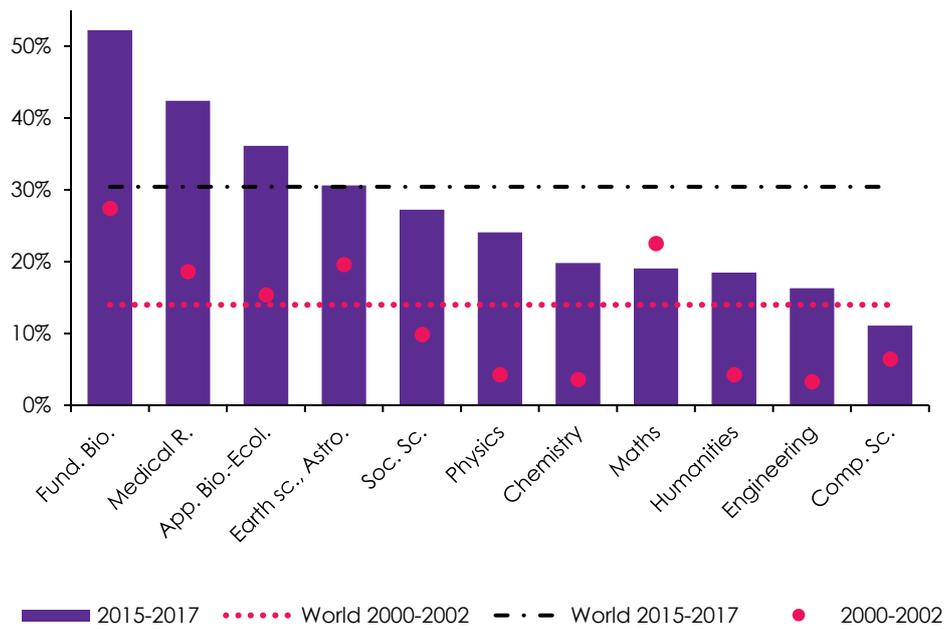

Source: Computed by author using OST-WoS database

With regard to the nomenclature in 11 disciplines of the OST, like the ERC sub-fields, the proportion of OA publications is relatively high in fundamental biology (52%) and in medical research (42%) and low in humanities (18%), engineering (16%) and computer science (11%).

**Figure 4: share of OA publications, top 20 countries (2015-2017)**

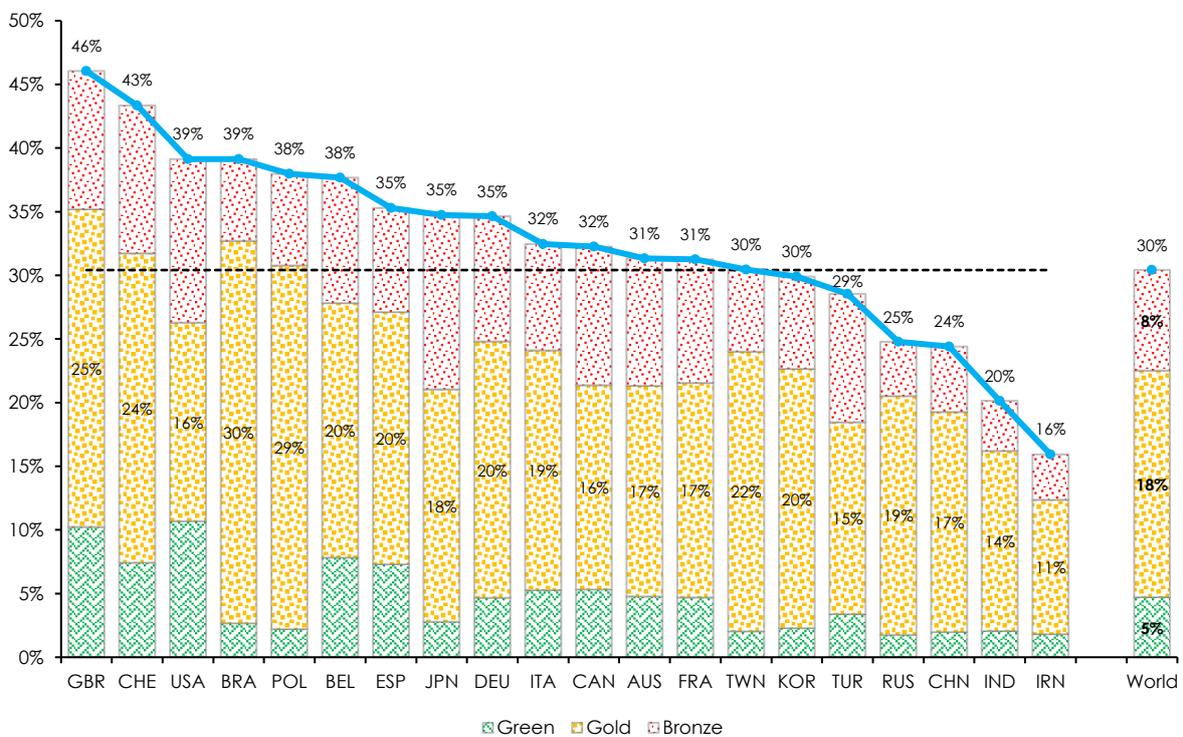

Source: Computed by author using OST-WoS database

Among the top 20 producers, the share of OA publications is contrasted. The United Kingdom is the country with the largest OA share (46%), 15% above the world average, followed by Switzerland (43%). The United States is followed by Brazil with a similar share of OA publications (39%). France has a slightly higher share than the world average (31%). Figure 4 also shows that countries with a high specialization engineering, mathematics and chemistry have low OA shares, such as Russia, China, India and Iran (OST, 2019).

In addition, the profile of the top 20 producing countries varies greatly depending on the type of OA. Although the share of OA publications is identical for Brazil and the United States, their profiles by type are very different. Most of Brazil's OA publications are in Gold type, while the United States has a fairly even profile between Gold, Green and Bronze. France, Australia and Canada have almost identical profiles very close to overall distribution of the world.

Overall, the country profiles are more or less close to the world average by type of OA. However, there are some important differences to note. For example, the practice of Green OA is much more common in the United States and the United Kingdom; it is twice as large (10% against 5% for the world).

## 4. Application of NOAI at country and institutional levels

In this section we calculate the share of OA and the normalized OA indicator (NOAI) over the period 2015-2017 for the top 50 producing countries and for 112 French university research institutions included in the IPERU program. Our objective is to show the difference in their positioning depending on whether we use OA share or the NOAI.

Figure 5 shows the rank of countries according to their share of OA publications (abscissa axis), and their rank according to the NOAI using normalization at OST disciplines level (ordinate axis). The rank is in descending order. That is, the countries with the highest OA share are to the right of axis. Rank "1" represents the country with the lowest share of OA. The figure shows that globally the two ranks are correlated. However, the rank changes considerably for some countries like Romania, which gains 21 places by normalizing the proportion of OA by discipline. Thus, the share of OA is equal to 29.3% for Romania, while the NOAI is 1.23, i.e. 23% higher than the world average (see annex 1). This is also the case of Ukraine (wins 19 places in the ranking) and Russia (15 places). This was expected because these countries are very specialized in engineering and/or computer science. We also note that a number of low-income countries specializing in "low-OA share disciplines" are moving up in ranking with the normalization.

**Figure 5: rank of top 50 producing countries by OA share and NOAI (normalization by OST disciplines), 2015-2017** – *bubbles size is proportional to the number of OA publications*

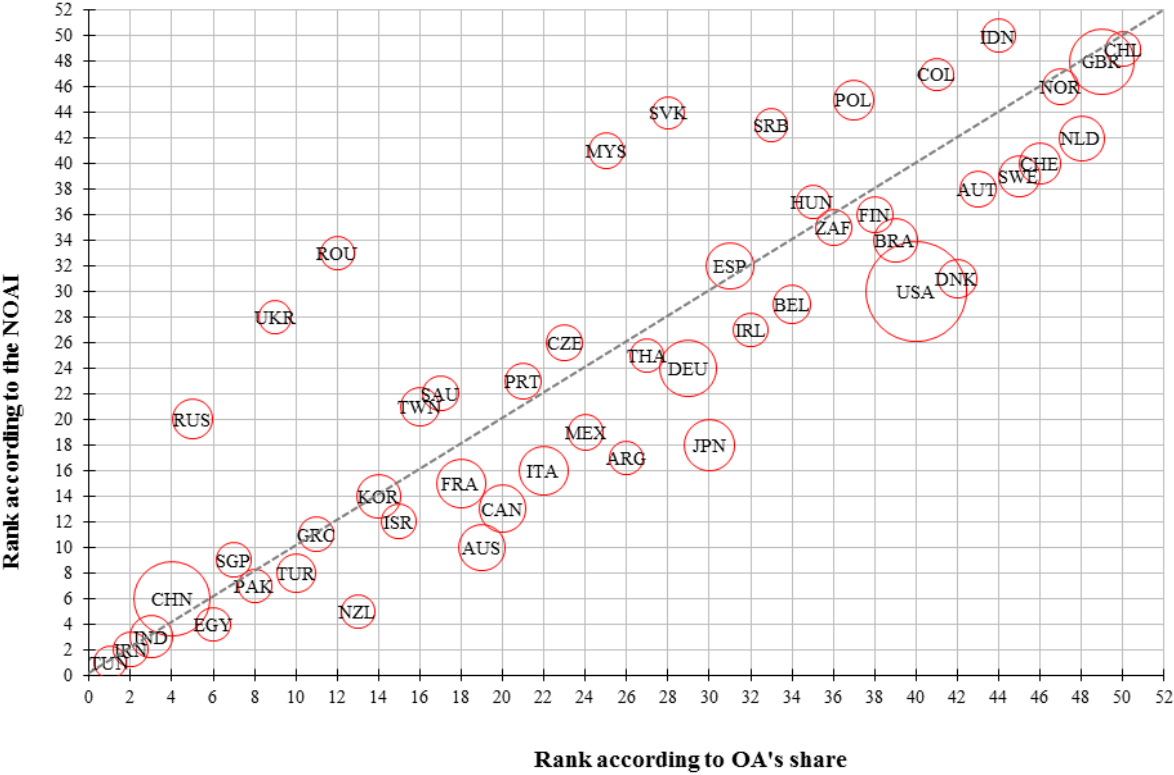

Source: Computed by author using OST-WoS database

Although France is very specialized in mathematics, it loses some places in the ranking on the normalized indicator. This could be explained by the fact that France has a relatively more diversified disciplinary profile, unlike low-income countries with disciplinary profiles that are very much oriented towards one or two disciplines. The United-Kingdom and Norway have the highest OA (share and NOAI) indicators compared to other high-income countries.

**Figure 6: rank of top 50 producing countries countries by OA share and NOAI (normalization by WoS subject categories), 2015-2017** – *bubbles size is proportional to the number of OA publications*

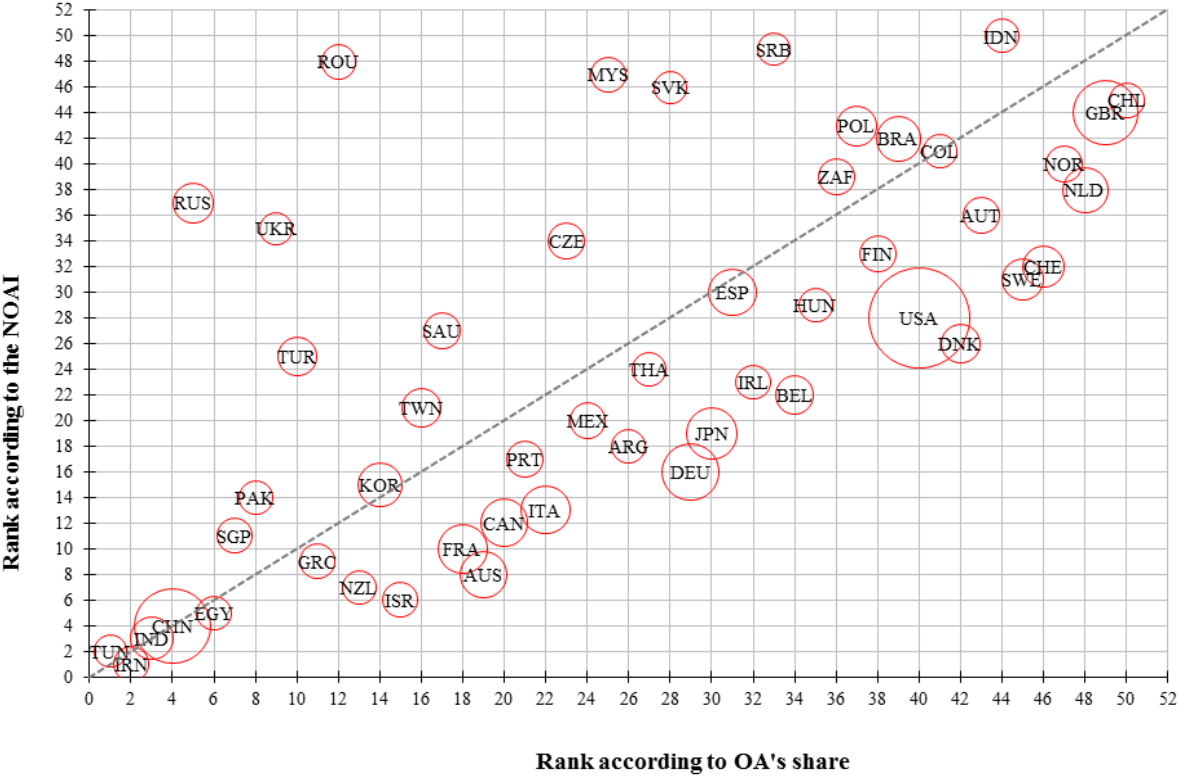

Source: Computed by author using OST-WoS database

The normalization at the level of the 254 WoS subject categories shows some differences compared to OST disciplines based normalization (Figure 6). Some countries keep their positions regardless of the method of normalization; while others change their positions like United-Kingdom losing 4 ranks and Turkey moving up by 17. Overall the ranks remain similar on both types of normalization. It should also be noted that the advantage obtained by normalization is more pronounced for several countries when we normalize by WoS subject categories, notably for Turkey (1.19 normalizing by WoS subject category versus 0.97 normalizing by OST discipline), Romania (1.57 versus 1.23), Ukraine and Russia (1.28 versus 1.16) (see annex 1).

**Figure 7: rank of French universitary institutions by OA share and NOAI (normalization by WoS subject categories), 2015-2017**

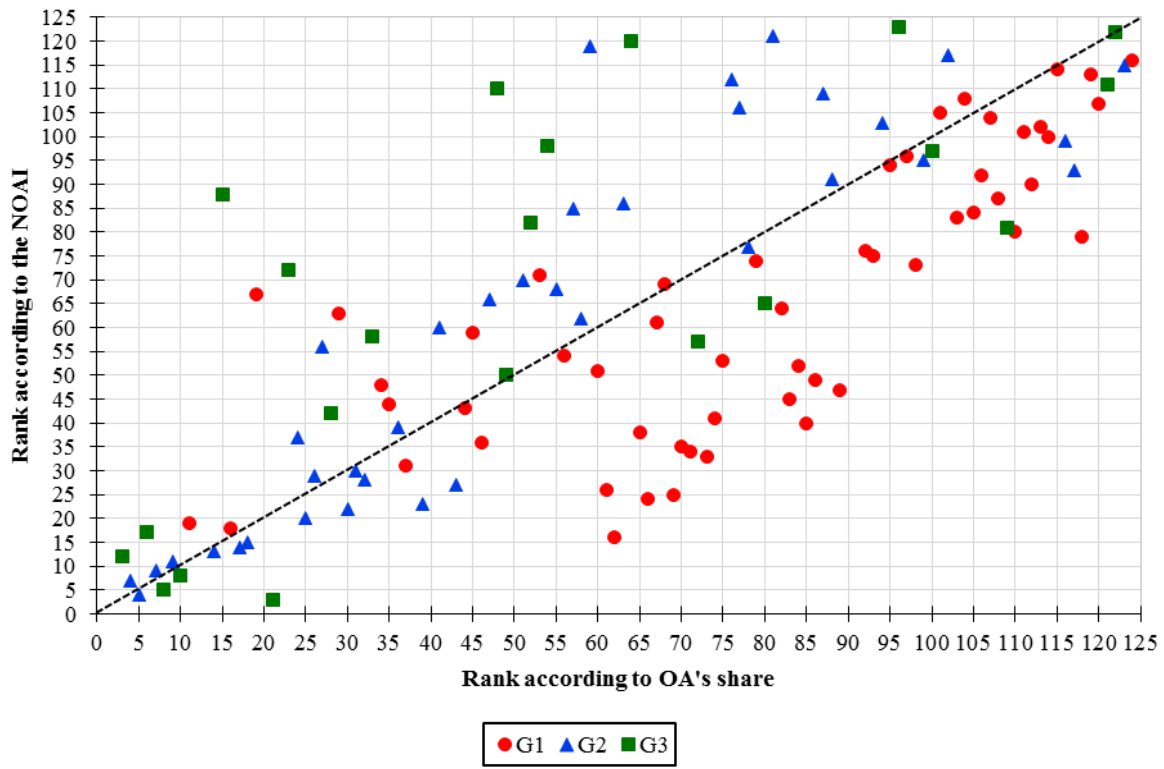

Source: Computed by author using OST-WoS database

Figure 7 shows that there is a large variation in ranking according to the indicator used, in particular for certain institutions[3]. The figure shows that a good part of schools of engineering and specialized institutions in engineering, computer science or social sciences and humanities are located to the left of the bisector (groups 2 and 3). That is, they gain rank after normalization of share of OA publications. In contrast, institutions to the right of the bisector are more oriented towards basic biology, applied biology-ecology, and medicine (group 1). The two rankings nevertheless remain globally correlated (the Spearman correlation coefficient is equal to 0.75).

---

[3] For confidentiality reasons, we cannot display the institution's names.

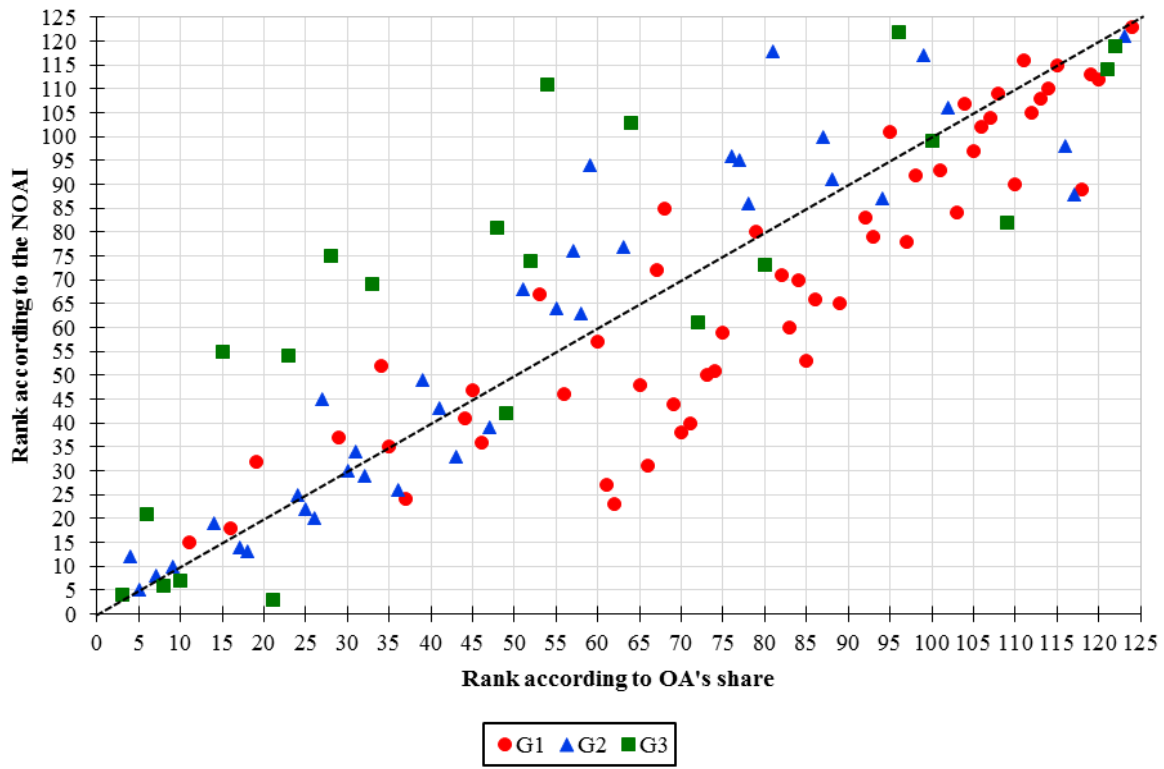

Figure 8: rank of French universitary institutions by OA share and NOAI (normalization by OST disciplines), 2015-2017

Source: Computed by author using OST-WoS database

Figure 8 shows that when normalization is carried out by OST disciplines, rank changes are relatively less important. The Spearman correlation coefficient of the two ranks (OA share and NOAI) is higher (0.85).

## 5. Conclusion and discussion

Through this paper, we have shown that OA publishing practices vary by discipline. The rate of "openness" is relatively high in the disciplines of life sciences such as basic biology or medicine. The rate is much lower in engineering or computer science.

In addition to disciplinary differences, there are several factors that can affect the measurement of OA. The perimeters considered can have a strong impact on the results obtained. Thus, the OA share depends on the OA's status type taken into account, the databases used, the restrictions made on the publications (taking into account or not publications without DOI), the type of publications included in the computation and the way to deal with disciplinary disparities.

The OA status type taken into account is an essential element in the computation of the share of OA publications. Not all scientific productions are published in journals. For example, the rate of publications in journals of deposits in the French open archive HAL is very low (around 3%). Integrating working papers or non-submitted articles deposited in open archives systematically increases the share of OA publications. Indeed, it is important to remember that the green OA status found in international databases such as WoS, only concerns books or articles published in journals, conference proceedings or books series. This should not be confused with invisible green OA in databases, which notably concerns articles not published in journals. For instance, an article in physics simply deposited in an open archive (like ArXiv) will be considered in the WoS data only if it has also been published in a journal indexed in the database. The "Green" status in databases only concerns articles deposited in archives AND published in journals indexed.

Restrictions on publications included in the perimeter also play an important role in determining the OA share. Especially the inclusion or not of publications that do not have a DOI. Scientific publication can only be freely accessed when it is published on the internet: all OA publications could then have a DOI. On the other hand, all publications with a DOI are not in OA. Restricting the perimeter to only publications having DOI could make sense. However, this choice, in fact, excludes all publications from "100% paper" journals (not OA therefore). The part of production that is not freely available will therefore be excluded from the denominator. This mechanically increases the OA share. In addition, it should be remembered that in the international databases such as WoS or Scopus, the completeness of the information relating to DOI varies from year to year. Information is satisfactory for recent years and much less so for older years (especially before 2010). This choice is restrictive for longitudinal studies. For these reasons, OST has chosen to work on the whole perimeter with or without DOI. Conversely, in CWTS calculations (also in their ranking), only publications with DOI are taken into account in the calculation of OA publications. Consequently, the OA share by institution is higher in CWTS productions.

Apart from the OA status type taken into account and the restrictions made on the perimeter (DOI or not), the publications type included in the study also impact the results. As it can be seen by comparing the results presented in the MESRI (2019) - French Ministry of Higher Education - and PSL (2019) – Paris Science et Lettres University - studies on the French case, OA is much more common for journal articles than for books or book chapters. Integrating them into calculation pulls the OA share down. In the PSL (2019) study, only journal articles are taken into account (with an OA share for France of 47% in 2017). Conversely, in the MESRI (2019) study, several types of publications are taken into account. In addition to journal articles, books, book chapters and conference proceedings are also integrated. OA share of the MESRI study was 41% in 2017, 6% lower than that of PSL. Since the perimeters are the same when it comes to the presence of DOI, the difference is explained in particular by the fact that the practice of OA is much more common for journal articles than for books or book chapters.

Notwithstanding the differences due to the chosen perimeters, the disciplinary structure of the actors is decisive when it comes to OA publications. In the literature, all studies dealing with OA agree that the share of OA vary considerably between disciplines. Consequently, an institution specialized in computer science will *de facto* have a lower OA rate than an institution whose specialization is in biology or even in medicine. Accounting for these disparities in institutions profiles seems essential for a correct comparison. The normalized indicator "NOAI" proposed in this paper consists in reporting at first the share of OA publications in a given discipline (or field) for an actor (institution, country, etc.) on the same share at the world level. In a second step, a weighted average of normalized OA shares by discipline (or for all disciplines) is calculated. These steps provide an overall indicator of OA corrected for disciplinary differences in terms of openness. Two levels of aggregation are used for normalization. A more general level of aggregation comprising 11 disciplines (for OST nomenclature see annex 2), and a fine aggregation level represented by the 254 subject categories of the WoS database. The results indicate that the normalization obtained using the second level allows for better accuracy.

For decision-makers and for funders it is imperative to have a good grasp of all the parameters that may affect the calculation of OA publications, especially when it comes to allocating funds or grants. Like all other bibliometric indicators, the OA indicators must be handled with caution given the number of parameters that can affect them. Normalization corrects the simple OA share and makes possible comparisons within the same population (database). Given the important role of archives in the area of OA, these normalized indicators should be accompanied by other measures (also normalized) based on other sources than international databases.

## 6. Limitations

The shares of OA do not necessarily reflect the true practices of the disciplines, but rather give an image of the WoS database. Normalization allows some correction. Otherwise, there needs to be more discussion about the type of counting that should be used: geographical fractioning does not conceptually make much sense for this indicator, while it would allow making sums over countries (or institutions).

# Appendices

## 1. OA share, NOAI and OA number of publications by country, 2015-2017

| Countries | OA share | NOAI (subject categories normalization) | NOAI (OST's disciplines normalization) | Number of OA publications |
|---|---|---|---|---|
| USA | 39,13 | 1,22 | 1,21 | 182981 |
| CHN | 24,40 | 0,94 | 0,92 | 92747 |
| GBR | 46,05 | 1,48 | 1,48 | 64357 |
| DEU | 34,64 | 1,11 | 1,13 | 44719 |
| JPN | 34,75 | 1,15 | 1,10 | 32251 |
| FRA | 31,26 | 1,03 | 1,04 | 28039 |
| ITA | 32,46 | 1,03 | 1,05 | 26887 |
| CAN | 32,27 | 1,03 | 1,02 | 25364 |
| ESP | 35,28 | 1,25 | 1,22 | 24553 |
| AUS | 31,33 | 1,00 | 0,98 | 22700 |
| NLD | 44,97 | 1,32 | 1,36 | 20741 |
| BRA | 39,13 | 1,44 | 1,25 | 20078 |
| KOR | 29,90 | 1,11 | 1,02 | 19467 |
| IND | 20,15 | 0,90 | 0,78 | 15912 |
| CHE | 43,35 | 1,27 | 1,35 | 15441 |
| SWE | 42,02 | 1,27 | 1,30 | 13471 |
| POL | 37,99 | 1,46 | 1,39 | 13093 |
| RUS | 24,77 | 1,31 | 1,10 | 12280 |
| TUR | 28,56 | 1,19 | 0,97 | 9855 |
| BEL | 37,68 | 1,17 | 1,18 | 9834 |
| TWN | 30,44 | 1,15 | 1,11 | 9226 |
| DNK | 40,32 | 1,20 | 1,22 | 8773 |
| AUT | 40,32 | 1,29 | 1,30 | 7639 |
| NOR | 43,65 | 1,38 | 1,39 | 6931 |
| FIN | 38,41 | 1,28 | 1,28 | 6030 |
| ZAF | 37,88 | 1,35 | 1,26 | 5913 |
| PRT | 32,30 | 1,14 | 1,12 | 5836 |
| CZE | 32,71 | 1,28 | 1,16 | 5700 |
| IRN | 15,93 | 0,77 | 0,63 | 5667 |
| MYS | 33,14 | 1,56 | 1,36 | 5588 |
| MEX | 33,01 | 1,15 | 1,10 | 5504 |
| ISR | 30,17 | 0,99 | 1,00 | 5015 |
| SAU | 30,69 | 1,20 | 1,12 | 4853 |
| CHL | 46,41 | 1,49 | 1,57 | 4560 |
| SGP | 26,09 | 1,03 | 0,98 | 4345 |
| GRC | 28,87 | 1,01 | 1,00 | 3912 |
| ROU | 29,33 | 1,57 | 1,23 | 3688 |
| IRL | 35,36 | 1,18 | 1,16 | 3521 |
| ARG | 33,85 | 1,14 | 1,07 | 3417 |
| THA | 33,89 | 1,18 | 1,13 | 3338 |
| NZL | 29,43 | 0,99 | 0,92 | 3289 |
| IDN | 41,98 | 2,07 | 1,79 | 3273 |
| EGY | 25,12 | 0,97 | 0,85 | 3180 |
| HUN | 37,87 | 1,25 | 1,29 | 3136 |
| PAK | 26,36 | 1,06 | 0,94 | 2871 |
| COL | 39,60 | 1,40 | 1,45 | 2233 |
| SRB | 36,77 | 1,61 | 1,39 | 2189 |
| SVK | 34,28 | 1,50 | 1,39 | 1859 |
| UKR | 26,88 | 1,28 | 1,16 | 1478 |
| TUN | 15,12 | 0,78 | 0,61 | 840 |

## 2. OST disciplines

| OST disciplines* | Abbreviations |
|---|---|
| Applied biology - Ecology | App. Bio. - Eco. |
| Fundamental biology | Fund. bio. |
| Chemistry | Chemistry |
| Computer science | Comp. Sc. |
| Mathematics | Maths |
| Physics | Physics |
| Medical research | Medical R. |
| Engineering | Engineering |
| Earth sciences – Astronomy – Astrophysics | Earth sc., Astro. |
| Humanities | Humanities |
| Social sciences | Soc. Sc. |

* OST disciplinary classification is available on: https://figshare.com/articles/OST_disciplinary_classification/11897601

## 3. ERC based classification

| ID | Wording |
|---|---|
| SH1 | Individuals, Markets and Organizations |
| SH2 | Institutions, Values, Environment and Space |
| SH3 | The Social World, Diversity, Population |
| SH4 | The Human Mind and Its Complexity |
| SH5 | Cultures and Cultural Production |
| SH6 | The Study of the Human Past |
| PE1 | Mathematics |
| PE2 | Fundamental Constituents of Matter |
| PE3 | Condensed Matter Physics |
| PE4 | Physical and Analytical Chemical Sciences |
| PE5 | Synthetic Chemistry and Materials |
| PE6 | Computer Science and Informatics |
| PE7 | Systems and Communication Engineering |
| PE8 | Products and Processes Engineering |
| PE9 | Universe Sciences |
| PE10 | Earth System Science |
| LS1 | Molecular Biology, Biochemistry, Structural Biology and Molecular Biophysics |
| LS2 | Genetics, 'Omics', Bioinformatics and Systems Biology |
| LS3 | Cellular and Developmental Biology |
| LS4 | Physiology, Pathophysiology and Endocrinology |
| LS5 | Neuroscience and Neural Disorders |
| LS6 | Immunity and Infection |
| LS7 | Applied Medical Technologies, Diagnostics, Therapies and Public Health |
| LS8 | Ecology, Evolution and Environmental Biology |
| LS9 | Applied Life Sciences, Biotechnology, and Molecular and Biosystems Engineering |

Source : https://erc.europa.eu/content/erc-panel-structure-2019